\documentstyle[10pt,psfig]{article}
\begin{document}
\begin{center}
{\bf Speckle Imaging: a boon for astronomical observations}
\end{center}
\vspace{0.5cm}

\begin{center}
{\bf S. K. Saha \\                           
Indian Institute of Astrophysics\\      
Bangalore - 560 034} \\                  
e-mail: sks@iiap.ernet.in\\
\end{center}
\vspace{0.5cm}

\begin{center}
{\bf Abstract}
\end{center}
\vspace{0.3cm}

\noindent
The speckle imaging is a photographic technique that resolves objects viewed
through severely distorted media. The results are insensitive to the errors
caused by apparent size of the isoplanatic patch and the telescope aberrations.
In this article, a short descriptions of the atmospheric turbulence
and its effect on the flat wavefront from a stellar source is presented; 
the shortcomings of the conventional long-exposure images in the presence of 
Earth's atmosphere are discussed. The advantages of the speckle interferometric 
technique over conventional imaging are enumerated. The technical details of 
the method, basic Fourier optics, data analysis procedures are also described.
\vspace{0.3cm}

\begin{center}
{\bf 1. Introduction}
\end{center}
\vspace{0.3cm}

\noindent
The atmosphere of earth restricts the resolution of conventional ground 
based astronomical images to about 1 arc second. This is due to the refractive
index variations of the atmosphere through which the light rays reaching the
telescope. When a star image is observed through a telescope with high 
magnification, the observed image structure is usually far from the theoretical 
pattern. The appearance of image depends strongly on the size of aperture of the 
telescope. Large telescope helps in gathering more optical energy, as well as
in obtaining better angular resolution. The resolution increases with the
diameter of the aperture. 
For example, the diffraction-limit of a 2.34 meter telescope is about 
0.05 arc second. But in reality, the image is degraded by factor of 20.
With small aperture a random motion of a image is often affect the main effect, 
whereas, with large aperture spreading and blurring of the image occur.
\vspace{0.3cm}

\noindent
Owing to the diffraction phenomenon, the image of
the point source (unresolved stars) cannot be smaller than a limit 
at the focal plane of the 
telescope. This phenomenon can be observed in ocean, when regular waves pass 
through an aperture. It is present in the sound waves, as well as in the
electro-magnetic spectrum too starting from gamma rays to radio waves.
\vspace{0.3cm}

\noindent
Speckle interferometric technique (Labeyrie, 1970) yields the 
diffraction-limited autocorrelation of the object. The diffraction-limited 
resolution of celestial objects viewed through the Earth's turbulent 
atmosphere could be achieved with the largest optical telescope, by post 
detection processing of a large data set of short-exposure images using 
Fourier-domain methods. Certain specialized moments of the Fourier transform of 
a short-exposure image contain diffraction-limited information about the object 
of interest. In this article, I shall describe a few basic theorems of Fourier 
optics, which are essential to understand this technique followed by the theory 
of speckle interferometry. The data processing method to analyze 
specklegrams of close binary stars obtained 
with 2.34 meter Vainu Bappu telescope (VBT), Kavalur, India is also discussed.
\vspace{0.3cm}

\begin{center}
{\bf 2. Preamble}
\end{center}
\vspace{0.3cm}

\noindent
In order to obtain high angular resolution of an stellar object, Fizeau (1868) 
had suggested to install a screen with two holes on top of the 
telescope that produce Young's fringes at its focal plane as the 
fringes remain visible in presence of seeing, thus
allowing measurements of stellar diameters.
Stefan attempted with 1 meter telescope at Observatoire de Marseille but could 
not notice any significant drop of fringe visibility and opined
none of the observed stars approached 0.1 arc-seconds in angular size. 
About half a century later, Michelson could measure the diameter of the 
satellites of Jupiter with Fizeau interferometer on top of the Yerkes refractor. 
With the 100 inch telescope at Mt. Wilson (Anderson, 1920),  
the angular separation of spectroscopic binary star Capella was measured.  
\vspace{0.3cm}

\noindent
To overcome the restrictions of the baseline, Michelson (1920) constructed the 
stellar interferometer equipped with 4 flat mirrors to fold the beams  
by installing a 7 meter steel beam on top of the afore-mentioned 100 
inch telescope; the supergiant star $\alpha$~Orionis were resolved  
(Michelson and Pease, 1921). Due to the various difficulties, viz., (i) effect 
of atmospheric turbulence, (ii) variations of 
refractive index above small sub-apertures of the interferometer, 
and (iii) mechanical instability, the project was abandoned.
\vspace{0.3cm}

\noindent
The field of optical interferometry lay dormant until it was revitalized by the
development of intensity interferometry (Brown and Twiss, 1958). Success in
completing the intensity interferometer at radio wavelengths (Brown et al.,
1952), in which the signals at the antennae are detected separately and the 
angular diameter of the source is obtained by measuring correlation of the
intensity fluctuations of the signals as a function of antenna separation,
Brown and Twiss (1958) demonstrated its potential at optical wavelengths by
measuring the angular diameter of Sirius. Subsequent development of this 
interferometer with a pair of 6.5 meter light collector on a circular railway
track spanning 188 meter (Brown et al., 1967), depicted the measurements of
32 southern binary stars with angular resolution limit of 0.5 milliarcseonds
(Brown, 1974). The project was abandoned due to lack of photons beyond 2.5
magnitude stars.
\vspace{0.3cm}

\noindent
Meanwhile, Labeyrie (1975) had developed a long baseline interferometer $-$ 
Interf$\grave{e}$rom$\bar{e}$tre $\grave{a}$ deux telescope (I2T) $-$ using a 
pair of 25~cm telescopes at Observatoire de Calern, France. His design 
combines features of the Michelson and of the radio interferometers. The use 
of independent telescopes increases the resolving capabilities. In this case, 
coude beams from both the telescopes arrive at central station and recombines 
them. This interferometer obtained the first measurements for a number of giant
stars (Labeyrie, 1985). Following the success of its operation, he undertook
a project of building large interferometer known as grand 
interf$\grave{e}$rom$\bar{e}$tre $\grave{a}$ deux telescope (GI2T) at the same
observatory. This interferometer comprises of two 1.5 meter spherical
telescopes on a North-South baseline, which are movable on a railway track
(Labeyrie et al., 1986). Mourard et al. (1989) had resolved the rotating
envelope of hot star $\gamma$~Cassiopeiae using this interferometer. The
technical details of this kind of interferometers can be found in the recent
article by Saha (1999a). 
\vspace{0.3cm}

\noindent
There are several long baseline interferometers that are in operation; some are 
at various stages of development (Saha, 1999a). These interferometers are based 
on the principle of merging speckles from both the telescopes. In other words, 
the fringed speckle can be visualized when a speckle
from one telescope is merged with the speckle from the other telescope. 
Therefore, it is necessary to get acquainted with relatively new topics,
speckle interferometry.
In what follows, the formation of speckles and the way to decode the 
atmospherically degraded informations are discussed in brief. 
\vspace{0.3cm}

\begin{center}
{\bf 3. Convolution and its applications}
\end{center}
\vspace{0.3cm}

\noindent
The convolution of two functions is a mathematical procedure (Goodman, 1968)
which simulates phenomena such as a blurring of a photograph. That may be 
caused by poor focus, by the motion of a photographer during the exposure, by 
dirt on the lens etc. In such blurred picture each point of object is 
replaced by a spread function. The spread function is disk shaped in the
case of poor focus, line shaped if the photograph has moved, halo shaped 
if there is a dust on lens. In other words, we know that delta function has 
value at a single point otherwise it is zero. But generally the measurement 
does not produce this. 
\vspace{0.3cm}

\noindent
Let us consider an input curve that can be represented by the 
curve $f(x)$ in terms of lot of close delta functions which are spread. Here, 
the shape of the response of the system including unwanted spread, is same 
for all values of $x$ (invariant for each considered delta function). Now, 
the value of the function $f(x)$ at $x_{1}$ is 
$f(x^{\prime}) \star g(x_{1}-x^{\prime})$. This is similar for each
considered point on the curve. So for the whole curve, we define mathematically

\begin{equation}
h(x) =  \int_{-\infty}^{+\infty} f(x^{\prime}) g(x-x^{\prime}) dx^{\prime},
\end{equation} 

\noindent
where, $h(x)$ is the output value at particular point $x$, $\star$ stands for
convolution. This integral is defined as convolution of $f(x)$ and $g(x)$. 

\begin{equation}
h(x) = f(x) \star g(x),
\end{equation} 

\noindent
where, $g(x)$ is referred to as a blurring function or line spread function 
(LSF) or in two dimensions, the point spread function (PSF).
\vspace{0.3cm}

\noindent
The Fourier transform of a convolution of two functions
is the product of the Fourier transform of the two functions. Therefore, in
the Fourier plane the effect becomes a multiplication, point by point, of
the transform of $\widehat{F}(u)$ with the transfer function $\widehat{G}(u)$.
\vspace{0.3cm}

\begin{center}
{\bf 4. Atmospheric turbulence and speckle formation}
\end{center}
\vspace{0.3cm}

\noindent
Owing to the turbulent phenomena associated with heat flow and winds in the
atmosphere, the density of air fluctuates in space and time. The  
inhomogeneities of the refractive index of the air can have devastating 
effect on the resolution achieved by any large telescope. The disturbance takes 
the form of distortion of the shape of the wavefront and variations of the 
intensity across the wavefront. Due to the motion
and temperature fluctuations in the air above the telescope aperture,
inhomogeneities in the refractive index develop. These inhomogeneities have
the effect of breaking the aperture into cells with different values of
refractive index that are moved by the wind across the telescope aperture.
\vspace{0.3cm}

\noindent
Kolmogrov law represents the distribution of turbule sizes, from millimeters 
to meters, with lifetimes varying from milliseconds to seconds. Changes in the 
refractive index in different portions of the aperture result to the phase 
changes in the value of the aperture function. The time evolution of the 
aperture function implies that the point spread function is time dependent. When 
the Reynolds number exceeds some value in a pipe (depending on the geometry of
the pipe), the transition from laminar flows to turbulent flows occur. The 
dimensionless quantity Reynolds number is defined as $QL/\nu$, where $Q$ is the 
mean flow speed, $L$ is the transverse size of the pipe and $\nu$ is kinematic
viscosity of the fluid. If $L$ is taken as some characteristic size of the
flow of atmosphere the result holds good for atmospheric case. 
\vspace{0.3cm}

\noindent
The power spectral density of refractive index fluctuations caused by the
atmospheric turbulence follows a power law with large eddies having greater
power (Tatarski, 1961). A plane wave propagating through the atmosphere of 
earth is distorted by refractive index variation in the atmosphere 
(troposphere); it suffers phase fluctuations and reaches the entrance 
pupil of a telescope with patches of random excursions in phase (Fried, 1966). 
Therefore, the image of the star in the focal plane of a large telescope is 
larger then the Airy disk of the telescope. The size is equivalent to the 
atmospheric point spread function (point spread function is a modulus square of 
the Fourier transform of the aperture function). The resolution at the image 
plane of the telescope is determined by the width of the PSF which is of the 
order of $(1.22\lambda/r_\circ)$, where, $\lambda$ is a wavelength of light 
and $r_\circ$ is the average size of the turbulence cell, which is of the 
order of 10~cm. Therefore, resolution ($\sim$~1 arc second) 
is unfortunately much larger then the theoretical size $1.22\lambda/D$ of the 
Airy disk of a large telescope (Rayleigh limit or diffraction limit), where, 
$D$ is the diameter of the telescope. 
\vspace{0.3cm}

\noindent
The variance of phase difference fluctuations between any two points in the 
wave-front increases as the 5/3 power of their separation. When this variance
exceeds $\pi^2$~rad for some separation $r_\circ$, then all details in the
smaller than $\lambda/r_\circ$ will be obliterated in the long exposure
images. If the exposure time is shorter than the evolution time of the phase
inhomogeneities, then each patch of the wave-front with diameter $r_\circ$
$-$ Fried parameter $-$ would act independently of the rest of the wavefront
resulting in multiple images of the source. These sub-images or `speckles', as 
they are called and spread over the area defined by the long exposure image,
can occur randomly along any direction within an angular patch of diameter 
$\lambda/r_\circ$. The average size of the speckle is of the same order of 
magnitude as the Airy disk of the telescope in the absence of atmospheric 
turbulence and the lifetime of individual speckle is of the order of 0.1 to 
0.01 seconds. Figure 1 depicts the speckles of the star
HR4689; observations were carried out at 2.34 meter VBT,
Kavalur, India with the speckle interferometer (Saha et al., 1997, 1999a). 
\vspace{0.3cm}

\noindent
\begin{figure}[h]
\centerline{\psfig{figure=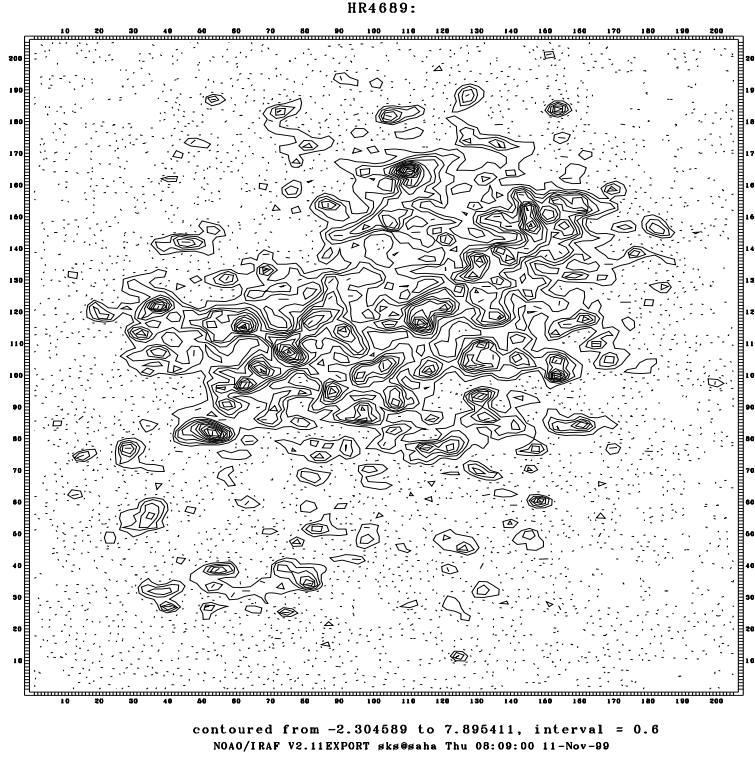,height=10cm,width=10cm,angle=270}}
\caption{Specklegram of a binary star, HR4689 obtained at VBT, Kavalur, India.
}
\end{figure}

\noindent
A snap shot taken later will show a different pattern but with similar 
probability of the angular distribution. A sum of similar exposures is the
conventional image. It is easy to visualize that the sum of several 
statistically uncorrelated speckle patterns from a point source can result in 
an uniform patch of light a few arc-seconds wide (Saha, 1999b). 
figure 2 shows the result of summing 128 specklegrams demonstrating the 
destructions of finer details of the image by the atmospheric turbulence.
\vspace{0.3cm}

\noindent
\begin{figure}[h]
\centerline{\psfig{figure=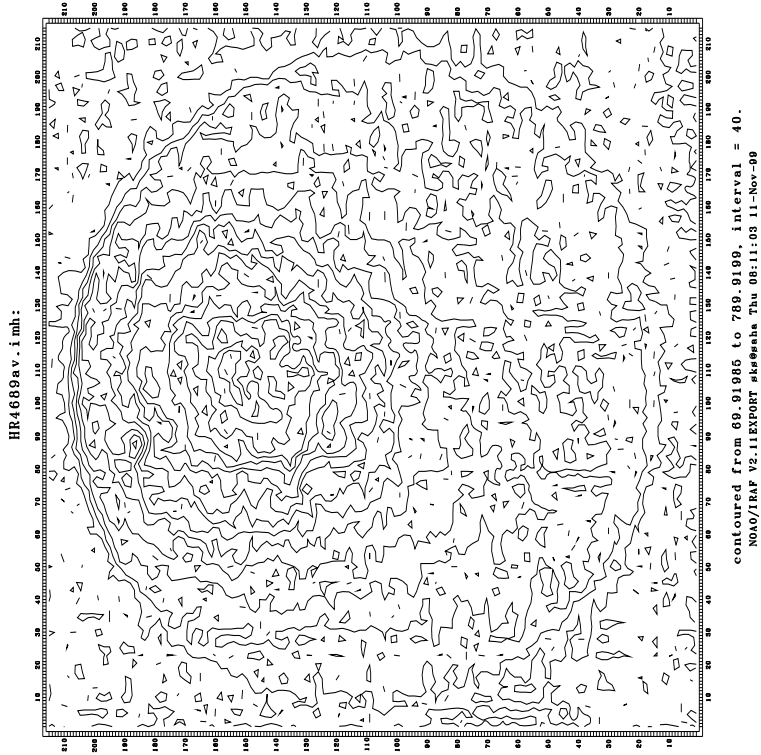,height=10cm,width=10cm}}
\caption{The result of summing 128 specklegrams of the same star, HR4689. 
}
\end{figure}

\noindent
Venkatakrishnan et al., (1989) had also generated the intensity distribution 
for $r_\circ$ in the plane of a large telescope; 
the smallest contours have the size of the Airy disk of the telescope.  
The result of summing 100 such distributions showed similar concentric 
circle of equal intensity, destroying the finer details of an image. The method
of such computer simulations runs as follows:
\vspace{0.3cm}

\noindent
The intensity distribution in the focal point of the telescope 
for the atmospheric cell ($r_\circ$) of 10cm size and $L_\circ$ = 200 cm for an 
entrance pupil of 200cm diameter were taken as samples. A power spectral 
density of the form

\begin{equation}
{\rho(k) {\propto} {\frac {L_\circ^{\frac{11}{3}}} {(1 + k^{2}L_\circ^{2}
)^{\frac {11}{6}}}}},
\end{equation}

\noindent
was multiplied with a random phase factor $e^{i\phi}$, one for each value
of $(k_x, k_y)$, with $\phi$ uniformly distributed between ${-\pi}$ and $\pi$.
\vspace{0.3cm}

\noindent
The resulting 2-D pattern in $k_x, k_y$ space was Fourier transformed
to obtain one realization of the wavefront $W(x, y)$. The Fraunhofer diffraction
pattern of a piece of this wavefront with the diameter of the entrance pupil
gives angular distribution of amplitudes, while the squared modulus of this 
field gives the intensity distribution in the focal plane of the telescope.
\vspace{0.3cm}

\noindent
The long-exposure PSF is defined by the ensemble average, $<S(x, y)>$,
independent of any direction. The average illumination, $I(x, y)$ of a
resolved object, $O(x, y)$ obeys convolution relationship,

\begin{equation}
<I(x, y)> = O(x, y) \star <S(x, y)>,
\end{equation}

\noindent
where, $(x, y)$ is a 2-dimensional space 
vector. Using 2-dimensional Fourier transform, this equation can be read as,

\begin{equation}
<\widehat{I}(u, v)> = \widehat{O}(u, v) \cdot <\widehat{S}(u, v)>,
\end{equation}

\noindent
where, $\widehat{O}(u, v)$ is the object spectrum, $<\widehat{S}(u, v)>$
is the transfer function for long-exposure images and is the product of the
transfer function of the atmosphere $\widehat{B}(u, v)$, as well as the 
transfer function of the telescope, $\widehat{T}(u, v)$. $(u, v)$ is the
spatial frequency vector. The transfer function for 
long-exposure image can be expressed as,

\begin{equation}
<\widehat{S}(u, v)> = \widehat{B}(u, v) \cdot \widehat{T}(u, v).
\end{equation}

\noindent
The benefit of the short-exposure images over long-exposure can be 
visualized by the following explanation.
\vspace{0.3cm}

\noindent
Let us consider two seeing cells separated by a vector in the telescope pupil, 
$\lambda (u, v)$, where $\lambda$ is the mean wavelength. 
If a point source is imaged through the 
telescope by using pupil function consisting of two apertures, corresponding
to the two seeing cells, then a fringe pattern is produced with narrow spatial 
frequency bandwidth. If the major component $\widehat{I}(u, v)$ at the 
frequency $(u, v)$ is produced by contributions from all pairs of points with 
separations $\lambda (u, v)$, with one point in each aperture and is averaged 
over many frames, then the result for frequencies greater than $r_o/\lambda$ 
tends to zero. The Fourier component performs a random walk in the complex
plane and average to zero, $<\widehat{I}(u, v)>$ = 0, when $u > r_o/\lambda$.
\vspace{0.3cm}

\noindent
For a large telescope, the aperture, P, can be sub-divided into a set of
sub-apertures, $p_i$. According to the diffraction theory (Born and Wolf, 1984)
the image at the focal plane of the telescope is obtained by adding all such 
fringe patterns produced by all possible pairs of sub-apertures. With increasing
distance of the baseline between two sub-apertures, the fringes move with an 
increasingly larger amplitude. On a long-exposure images, no such shift is
observed, which implies the loss of high frequency components of the image.
While, in the short-exposure images ($<$20~msec), the interference fringes are 
preserved. 
\vspace{0.3cm}

\noindent
{\bf 4.1 Seeing }
\vspace{0.3cm}

\noindent
The common query for an observer is about `seeing'. This important parameter 
changes very fast, every now and then. It is the total effect of distortion in 
the path of the light via different contributing layers of the atmosphere 
to the detector placed at the focus of the telescope (detail discussions can 
be found in recent article by Saha, 1999a). The major sources of image 
degradation predominantly comes from the surface layer, as well as from 
the aero-dynamical disturbances in the atmosphere surrounding the 
telescope and its enclosure, namely, (i) thermal distortion of primary and 
secondary mirrors when they get heated up, (ii) dissipation of heat by the
latter mirror, (iii) rise in temperature at the primary cell, (iv) at the focal
point causing temperature gradient close to the detector etc. Saha and 
Chinnappan (1999) have found that the seeing at VBT improved gradually in the 
latter part of the night. 
\vspace{0.3cm}

\noindent
The resolution $\theta$ of a large telescope, limited by the atmospheric
turbulence, as defined by the Strehl criterion is 

\begin{equation}
{\theta = {\frac {4}{\phi}} {\frac {\lambda}{r_{o}}}},
\end{equation}

\noindent
Then the question arises how to measure seeing? The qualitative method is from 
the short exposure images using speckle interferometric technique, where,
the area of the telescope aperture divided by the estimated
number of speckles gives the wavefront coherence area $\sigma$, from which
$r_{o}$ can be found by using relation, 

\begin{equation}
{\sigma = {0.342 {\left( \frac {r_{o}} {\lambda}\right) }^{2}}}.
\end{equation}

\noindent
If the autocorrelations of the short exposure images are summed, it contains  
autocorrelation of the seeing disk together with the autocorrelation of the 
mean speckle cell. It is width of the speckle component of the 
autocorrelation that provides information on the size of the object being 
observed (Saha and Chinnappan, 1999, Saha et al., 1999a). 
\vspace{0.3cm}

\noindent
Systematic studies 
of this parameter would enable to understand the various causes of the local 
seeing, for example, thermal inhomogeneities associated with the building, 
aberrations in the design, manufacture and alignment of the optical train etc. 
\vspace{0.3cm}

\noindent
Doom seeing plays a vital role in deteriorating image 
quality. It is necessary to take precautionary measure to avoid hot air 
entrapment. The important ones are (i) bring down the control room from the
observing floor, (ii) improve the cross ventilation at each of the floors, (iii)
remove unused machines or any other equipments, excess cemented portions
available around the building, as well as all glass enclosures from the 
building.  
\vspace{0.3cm}

\noindent
Mirror seeing is another important source of image spread and has the longest 
time-constant. The spread amounts to 0.5" for a 1$^\circ$ different in 
temperature. Therefore, It is essential to make arrangement to cool the 
primary mirror and try to maintain uniform temperature in and around
the primary mirror cell (Saha and Chinnappan, 1999).
\vspace{0.3cm}

\begin{center}
{\bf 5. Speckle}
\end{center}
\vspace{0.3cm}

\noindent
The term 'Speckle' refers to a grainy structure observed when an uneven surface
of an object is illuminated by a fairly coherent source. A good example of
speckle phenomena may be observed at the river port when many boats are
approaching towards the former at a particular time or in the swimming pool
when many swimmers are present. Each boat or swimmer emits wave trains and
interference between these random trains causes a speckled wave field on the
water surface. Depending on the randomness of the source, spatial or
temporal, speckles tend to appear. Spatial speckles may be observed when all
parts of the source vibrate at same constant frequency but with different
amplitude and phase, while temporal speckles are produced if all parts of it
have uniform amplitude and phase. With a non-monochromatic vibration spectrum,
in the case of random sources of light, spatio-temporal speckles are produced.
\vspace{0.3cm}

\noindent
The ground illumination produced by any star has fluctuating speckles, known
as star speckles. It is too fast and faint, therefore, cannot be seen directly.
Atmospheric speckles can be observed easily in a star image at the focus of a 
large telescope using a strong eyepiece. The star image looks like a pan of
boiling water. If a short exposure image is taken, speckles can be recorded.
The number of correlation cells
is determined by the equation $N = D/r_{o}$. As the seeing improves, the 
number decreases. It is clear that speckles are caused by interference effect 
between wave element having random phases, rather then by ray bending effect.   
Its structure in astronomical images is the result of 
constructive and destructive bi-dimensional interferences between rays coming 
from different zones of incident wave. The statistical properties of speckle 
pattern depend both on the coherence of the incident light and the properties 
of random medium.   
\vspace{0.3cm}

\noindent
Mathematically, speckles are simply the result of adding many sine functions 
having different, random characteristics. Since the positive and negative 
values cannot cancel out everywhere, adding an infinite number of such sine 
functions would result in a function with 100$\%$ constructed oscillations.
\vspace{0.3cm}

\noindent
{\bf 5.1 Imaging in the Presence of Atmosphere}
\vspace{0.3cm}

\noindent
In the ideal condition, the resolution can be achieved in an imaging experiment,
is limited only on imperfections in the optical system. If a collimated beam 
passes through the atmosphere and is collected by a telescope, the quality of 
the image formed is influenced by the atmospherically produced disturbances. 
\vspace{0.3cm}

\noindent
Let the modulation transfer function (MTF) of an optical system be
composed of the atmosphere and a telescope. The random wavefront tilt displaces
the image but does not reduce its sharpness. If a short exposure is recorded,
the image sharpness and MTF are insensitive to the tilt. While, in the case of
long exposure ($>$ 20 msec) image, the image is spread during the exposure
by its random variations of the tilt. Therefore, the image sharpness and the
MTF are affected by wavefront tilt, as well as by the more complex shapes. In
the short exposure case, a random factor associated with the tilt is extracted
from the MTF before being taken the average, where in the long exposure case,
no such factor is removed.
\vspace{0.3cm}

\noindent
Let us consider an imaging system consists of a simple lens based telescope
in which the point spread function (PSF) is invariant to spatial shifts. An
object (point source) at a point $(x^\prime, y^\prime)$ anywhere in the field of
view will, therefore, produce a pattern $S[(x, y) - (x^\prime, y^\prime)]$ 
across the image. If the object can emit incoherently, the image $I(x, y)$ of a 
resolved object $O(x, y)$ obeys a convolution relationships. The 
mathematical description of the convolution of two functions is of the form:

\begin{equation}
I(x, y) = \int O(x^\prime, y^\prime) S[(x, y) - (x^\prime, y^\prime)] 
d(x^\prime, y^\prime),
\end{equation}
\vspace{0.3cm}

\noindent
{\bf 5.2. Outline of the theory of speckle interferometry}
\vspace{0.3cm}

\noindent
By integrating autocorrelation function of the successive short-exposure 
records rather than adding the images themselves, the diffraction-limited 
information can be obtained. Indeed, autocorrelation function of a speckle 
images preserves some of the information in the way which is not degraded by 
the co-adding procedure. For each of the short-exposure instantaneous record, 
the quasi-monochromatic incoherent imaging equation applies,

\begin{equation}
I(x, y) = O(x, y) \star S(x, y),
\end{equation}

\noindent
where, $I(x, y)$ is the instantaneous image intensity, $O(x, y)$ the object 
intensity, $S(x, y)$ the instantaneous PSF.  
\vspace{0.3cm}

\noindent
The analysis of data may be carried out in two equivalent ways.
In the spatial domain the ensemble average space autocorrelation is found 
giving the resultant imaging equation.
\vspace{0.3cm}

\noindent
In the Fourier plane the effect becomes a multiplication, point by point, of
the transform of the object $\widehat{O}(u, v)$ with the transfer function
$\widehat{S}(u, v)$, and therefore, equation 10 leads to,

\begin{equation}
\widehat{I}(u, v) = \widehat{O}(u, v) \cdot \widehat{S}(u, v).
\end{equation}

\noindent
The ensemble average of the power spectrum is given by,

\begin{equation}
\ < |\widehat{I}(u, v)|^{2} \ > = |\widehat{O}(u, v)|^{2} \cdot \ 
< |\widehat{S}(u, v)|^{2} \ >.
\end{equation}

\noindent
Hence, if $\ < |\widehat{S}(u, v)|^{2}\ >$ is known, 
$|\widehat{O}(u, v)|$$^{2}$ can be 
estimated. To find this, one has to observe a point source close
to the object. The Fourier transform of a point source (delta function)
is a constant, $C_{n}$. Then, equation 12, for a point source is 

\begin{equation}
\ < |\widehat{I}_{s}(u, v)|^{2} \ > = C_{n}^{2} \cdot \ 
< |\widehat{S}(u, v)|^{2} \ >.
\end{equation}

\noindent
To find $C_{n}$$^{2}$,  one has to find the boundary condition. At the
origin of the Fourier plane, $S(u=0, v=0)$ is unity.  This is true for
an incoherent source.  Hence, $C_{n}$$^{2}$ is given by,

\begin{equation}
C_{n}^{2} = \ < |\widehat{I}_{s}(0, 0)|^{2} > / <|\widehat{S}(0, 0)|^{2} \ >,
\end{equation}

\noindent
i.e.,

\begin{equation}
C_{n}^{2} = \ < |\widehat{I}_{s}(0, 0)|^{2} \ >. 
\end{equation}

\noindent
Using equation 15 in equation 13 gives,

\begin{equation}
\ < |\widehat{S}(u, v)|^{2} \ > = \ < |\widehat{I}_{s}(u, v)|^{2} \ > /
\ < |\widehat{I}_{s}(0, 0)|^{2} \ >.
\end{equation}

\noindent
From equations 16 and 11, the power spectrum of the object is given by

\begin{equation}
 |\widehat{O}(u, v)|^{2} = \ < |\widehat{I}(u, v)|^{2} \ > /
[\ < |\widehat{I}_{s}(u, v)|^{2} \ > /\ < |\widehat{I}_{s}(0, 0)|^{2} \ ],
\end{equation}

\noindent
i.e., the power spectrum of the object is the ratio of the average power
spectrum of the image to the normalized average power spectrum of the point
source. By Wiener-Kinchin theorem, the inverse Fourier transform of equation
17 gives the autocorrelation of the object.

\begin{equation}
 A[O(x, y)] = FT^{-} [|\widehat{O}(u, v)|^{2}],
\end{equation}

\noindent
where, A stands for autocorrelation.
\vspace{0.3cm}

\begin{center}
{\bf 6. Observational technique at VBT}
\end{center}
\vspace{0.3cm}

\noindent
The programme of observing close binary systems (separation $<$ 1$\prime\prime$) 
has been going on since 1996 using speckle interferometer at the Cassegrain 
focus of the 2.34 VBT, Kavalur, India (Saha, 1999a). The details of this 
interferometer can be found in the articles (Saha et al., 1997, 1999a), that 
samples the image scale at the Cassegrain focus of the said telescope to
0.015" per pixel of the intensified CCD. The wave-front falls on the focal plane
and passes on to a microscope objective through a circular aperture of 
$\sim$350~$\mu$m of an optical flat kept at an angle of 15$^\circ$. This 
aperture was developed on a low expansion optical glass by devising a fine 
grinding precision mechanism at the laboratory (A. P. Jayarajan and S. K. Saha). 
The aperture was ground into the glass held at the angle of 75$^\circ$
with respect to the grinding axis. Interested readers may try it out to repeat 
the same. 
\vspace{0.3cm}

\noindent
The enlarged beam is recorded after passing through a
narrow band filter by a Peltier-cooled ICCD (386$\times$576) camera as 
detector which offers various option of exposure time, viz., 1~msecs, 5~msecs,
10~msecs, 20~msecs etc. It can operate in full frame, frame transfer and kinetic
modes. Since CCD is cooled to $\sim$40$^\circ$, the dark noise is considerably
low. Unlike the uncooled ICCD where data is stored in 8 bits, in this 
system, data is stored to 12 bits and can be archived to a Pentium PC. In
full frame, as well as in frame transfer modes, the region of interests can be
acquired at a faster speed. While in the kinetic mode, the image area can be
kept small to satisfy requirements, therefore, the rest of the area is usable 
for the data storage. The surrounding star field of diameter $\sim\phi$~10~mm gets
reflected from the optical flat on to a plane mirror and is re-imaged on to 
an uncooled ICCD (Chinnappan et al., 1991) for guiding the object.
\vspace{0.3cm}

\begin{center}
{\bf 7. Summary}      
\end{center}
\vspace{0.3cm}

\noindent
Among others the most important observations made by means of speckle 
interferometry is the discovery of compact cluster, R136a (HD38268), of 
Doradus nebula in the Large Magallanic Clouds (Weigelt and Baier, 1985).
Recent observations with adaptive optics system (Brandl et al., 1996) have
revealed over 500 stars within the field of view 12.8"$\times$12.8" covering
a magnitude range 11.2. Baba et al., (1994) have observed a binary star,
$\phi~And$ (separation 0.53") using imaging speckle spectroscopic method and
found that the primary star (Be star) has an H$\alpha$ emission line while the
companion has an H$\alpha$ absorption line.
\vspace{0.3cm}

\noindent
Developments of high angular resolution imaging have been going on in our
Institute over a decade. Several experiments have been conducted at Vainu
Bappu Observatory (Saha et al., 1987). The programme of speckle imaging at VBT
has been a successful one. Now we are in a position to obtain informations of
Fourier phase of the objects too (Saha et al., 1999b). Mapping of the certain 
interesting objects, viz., active galactic 
nuclei, proto-planetary nebulae will be undertaken in near future.
\vspace{0.3cm}

\begin{center}
{\bf References}      
\end{center}
\vspace{0.3cm}

\noindent
Anderson J. A., 1920, Astrophys. J., {\bf 51}, 263.

\noindent
Baba N., Kuwamura S., Miura N., Norimoto Y., 1994, Ap. J., {\bf 431}, L111.

\noindent
Born M., Wolf E., 1984, Principles of Optics, Pergamon Press.

\noindent
Brandl B., Sams B. J., Bertoldi F., Eckart A., Genzel R., Drapatz S., Hofmann R.,
Lowe M., Quirrenbach A., 1996, Ap. J., {\bf 466}, 254.

\noindent
Brown R. H., 1974, `The Intensity Interferometry, its Applications to Astronomy',
Taylor \& Francis, London.

\noindent
Brown R. H. and Twiss R. Q., 1958, Proc. Roy. Soc. A., {\bf 248}, 222.

\noindent
Brown R. H., Davis J. and Allen L. R., 1967, MNRAS, {\bf 137}, 375.

\noindent
Brown R. H., Jennison R. C. and Das Gupta M. K., 1952, Nature, {\bf 170}, 1061.

\noindent
Chinnappan V., Saha S. K., Faseehana, 1991, Kod. Obs. Bull. {\bf 11}, 87.

\noindent
Fizeau H., 1868, C. R. Acad. Sci. Paris, {\bf 66}, 934.

\noindent
Fried D. C., 1966, J. Opt. Soc. Am., {\bf 56}, 1972. 

\noindent
Goodman J. W., 1968, Introduction to Fourier optics, McGraw Hill Book Co. NY.

\noindent
Labeyrie A., 1970, Astron. \& Astrophys., {\bf 6}, 85.

\noindent
Labeyrie A., 1975, Astrophys. J., {\bf 196}, L71.

\noindent
Labeyrie A., 1985, 15th. Advanced Course, Swiss Society of Astrophys. and 
Astron. ed.. A. Benz, M. Huber and M. Mayor, 170.

\noindent
Labeyrie A., Schumacher G., Dugu$\acute{e}$ M., Thom C.,  Bourlon P., Foy F.,
Bonneau D. and Foy R., 1986, Astron. \& Astrophys., {\bf 162}, 359.

\noindent
Michelson A. A., 1920, Astrophys. J., {\bf 51}, 257.

\noindent
Michelson A. A., and Pease F. G., 1921, Astrophys. J., {\bf 53}, 249.

\noindent
Mourard D., Bosc I., Labeyrie A., Koechlin A. and Saha S., 1989, Nature,
{\bf 342}, 520.

\noindent
Saha S. K., 1999a, Bull. Astron. Soc. Ind., {\bf 27}, 443.

\noindent
Saha S. K., 1999b, Ind. J. Phys., {\bf 73b}, 553.

\noindent
Saha S. K., Chinnappan V., 1999, Bull. Astron. Soc. Ind., {\bf 27}, 327.

\noindent
Saha S. K., Jayarajan A. P., Sudheendra G., Umesh Chandra A., 1997, 
Bull. Astron. Soc. Ind., {\bf 25}, 379.

\noindent
Saha S. K., Sridharan R., Sankarasubramanian K., 1999b, submitted to 
Bull. Astron. Soc. Ind. 

\noindent
Saha S. K., Sudheendra G., Umesh Chandra A., Chinnappan V., 1999a, Exp. Astron,
{\bf 9}, 39. 

\noindent
Saha S. K., Venkatakrishnan P., Jayarajan A. P., Jayavel N., 1987, Curr. Sci.,
{\bf 56}, 985.
 
\noindent
Tatarski V. I., 1961, `Wave Propagation in a Turbulent Medium' McGraw Hill.

\noindent
Venkatakrishnan P., Saha S. K., Shevgaonkar R. K., 1989, Proc. `Image
Processing in Astronomy', ed., T. Velusamy, 57.

\noindent
Weigelt G., Baier G., 1985, A \& A., {\bf 150}, L18.
\vspace{1cm}

\begin{center}
{\bf Appendix I}
\end{center}
\vspace{1cm}

\begin{center}
{\bf Theorems of Fourier Transform}
\end{center}
\vspace{0.3cm}

\noindent
{\bf 1. Linearity theorem}
\vspace{0.3cm}

\noindent
\begin{equation}
F(\alpha g + \beta h) = \alpha F(g) + \beta F(h)
\end{equation} 

\noindent
i.e., the transform of sum of two functions is simply the sum of their
individual transforms.
\vspace{0.3cm}

\noindent
{\bf 2. Similarity theorem}
\vspace{0.3cm}

\noindent
 A stretching of the co-ordinates in the space domain $(x, y)$ results in the 
construction of the co-ordinates in the frequency domain $(f_x,~f_y)$ plus
a change in the overall amplitude of the spectrum. i.e., if 

\begin{equation}
F(g(x, y)) = G(f_{x}, f_{y})
\end{equation} 

\noindent
then, 

\begin{equation}
F(g(ax+by)) = G(f_{x}/a, f_{y}/b) /|ab|
\end{equation} 
\vspace{0.3cm}

\noindent
{\bf 3. Shift theorem}
\vspace{0.3cm}

\noindent
Translation of a function in a space domain introduces a linear phase
shift in the frequency domain. i.e., if

\begin{equation}
F(g(x, y)) = G(f_{x}, f_{y})
\end{equation} 

\noindent
then, 

\begin{equation}
F(g(x-a, y-b)) = G(f_{x}, f_{y}) \exp (-2\pi j (f_{x}a+f_{y}b))
\end{equation} 
\vspace{0.3cm}

\noindent
{\bf 4. Perseval's theorem}
\vspace{0.3cm}

\noindent
This theorem is generally interpretable as a statement of conservation of 
energy. It says that the total energy in the real domain is equal to the total
energy in the Fourier domain. i.e., 

\begin{equation}
F(g(x, y)) = G(f_{x}, f_{y})
\end{equation} 

\noindent
then

\begin{equation}
\int_{-\infty}^{+\infty} \int_{-\infty}^{+\infty} |g(x, y)|^{2}dxdy = 
\int_{-\infty}^{\infty} \int_{-\infty}^{\infty} 
|G(f_{x},f_{y})|^{2}df_{x} df_{y} 
\end{equation} 
\vspace{0.3cm}

\noindent
{\bf 5. Convolution Theorem}
\vspace{0.3cm}

\noindent
The convolution of two functions in the space domain (an operation that will be found to arise frequently in the
theory of linear system) is entirely equivalent of the more simple operation of multiplying their individual
transform. i.e., 

\begin{equation}
F(g(x, y)) = G(f_{x}, f_{y})
\end{equation} 

\noindent
and 

\begin{equation}
F(h(x, y)) = H(f_{x}, f_{y})
\end{equation} 

\noindent
then

\begin{equation}
F(\int_{-\infty}^{+\infty} \int_{-\infty}^{+\infty} 
g(\xi,\eta)h(x-\xi,y-\eta)d\xi d\eta) =  G(f_{x},f_{y})H(f_{x},f_{y})
\end{equation} 
\vspace{0.3cm}

\noindent
{\bf 6. Autocorrelation theorem}
\vspace{0.3cm}

\noindent
This theorem may be regarded as special case of convolution theorem. The 
Fourier transform of autocorrelation of
a function is the squared modulus of the Fourier transform. i.e. if 

\begin{equation}
F(g(x, y)) = G(f_{x}, f_{y})
\end{equation} 

\noindent
then, 

\begin{equation}
F(\int_{-\infty}^{+\infty} \int_{-\infty}^{+\infty}g(\xi,\eta)
g^{*}(\xi-x,\eta-y) d\xi d\eta) = |G(f_{x},f_{y})|^{2}
\end{equation} 
\end{document}